\documentclass[useAMS,usenatbib]{mn2e}

\usepackage{graphicx}
\usepackage{epsf}
\usepackage{psfig}

\title[An X-ray view of the INTEGRAL/IBIS blazars]{An X-ray view of the INTEGRAL/IBIS blazars}
\author[]{S. Giann\'i$^{1,2}$\thanks{E-mail:... }, A. De
Rosa$^{2}$, L. Bassani$^{3}$, A. Bazzano$^{2}$, T. Dean$^{4}$ and P. Ubertini$^{2}$\\
$^{1}$Universit\'a di Roma Tor Vergata\\
$^{2}$INAF-IASF Roma\\
$^{3}$INAF-IASF Bologna\\
$^{4}$School of Physics and Astronomy, University of Southampton,
SO17 1BJ, UK}

\begin{document}

\date{Accepted 2009 ... . Received 2009 ... ; in original form ...}

\pagerange{\pageref{firstpage}--\pageref{lastpage}} \pubyear{2002}

\maketitle

\label{firstpage}

\begin{abstract}

Aim of this work is a broad-band study with INTEGRAL, Swift and
XMM-Newton satellites of a sample of 9 blazars (7 FSRQ and 2 BL
Lac) with redshift up to about 4.

The spectral analysis has shown clear evidence of a flattening of
the continuum towards the low energies ($E<3$ keV observer frame).
This behaviour is well reproduced both with an absorbed power-law
model ($N_H\sim10^{20}$-$10^{23}$ cm$^{-2}$ in the rest-frame of
the sources) or a broken power-law continuum model (with an energy
break below $3$ keV in the observer-frame). No Compton reflection
features, Fe $K\alpha$ line and hump at high energies, have been
detected, with the exception of the source IGR J22517+2218 that
shows the presence of a weak iron line.

In this work we also investigate a possible correlation between
the absorption column density $N_H$ and the red-shift.

We confirm the existence of a $N_H$-z trend, with the higher
absorption at z$>$2 for a larger sample compared to previous
results. The distribution of the $N_H$ and the photon index
$\Gamma$ is also presented. The hard X-ray data allow us to detect
highly absorbed sources (with $N_H\ge10^{23}$cm$^{-2}$ in
rest-frame of the source) characterized by photon index
distribution peaked at harder values ($\Gamma\sim1.4$) with
respect to that obtained with XMM data only ($\Gamma\sim2$).


\end{abstract}


\section{Introduction}

Blazars are one of the most intriguing class of objects among the
Active Galactic Nuclei (AGN) and include BL Lacertae objects (BL
Lac) and Flat-Spectrum Radio Quasars (FSRQ). The intrinsic
differences between these two types of sources are in their
optical spectra and in the intensity of their emission: a BL Lac
source does not show any optical emission lines ($EW<5 \AA$) and
it is basically a low-power blazar ($L_{Bol}\sim10^{46}-10^{47}$
erg s$^{-1}$), whereas a FSRQ source shows significant emission
line equivalent widths and corresponds to a high-power blazar
($L_{Bol}\sim10^{48}$ erg s$^{-1}$) \citep{b60,b61}. In the AGN
unified scheme \citep{b59}, blazars are interpreted as
radio-sources with a relativistic jet aligned along the line of
sight \citep{b48}; in other words, the class of blazars is
represented by Radio Loud (RL) AGN observed very close to the
direction of the relativistic jets ($<10^o$).

The Spectral Energy Distribution (SED) of blazars is usually
modelled with two large humps produced via Synchrotron (peaking in
the infra-red to soft X-ray energy band) and Inverse Compton (IC)
emission, dominating the hard X-ray to gamma-ray regimes,
respectively. The lower energy emission component is Synchrotron
radiation from the jet whereas the high-energy component arises
through the inverse Compton of soft photons by highly relativistic
electrons in the jet plasma. These soft photons originate from the
local Synchrotron radiation within the jet or the nuclear
optical/UV emission, namely Synchrotron Self-Compton (SSC) and
External Compton (EC) components, respectively
\citep[see][]{b32,b9,b43}.

The X-ray spectrum of an AGN is usually well described by a
power-law with a specific flux (i.e. per unit energy interval) of
the form $N(E)\propto{E^{-\Gamma}}$, where E is the energy, N(E)
is the number of photons in units of s$^{-1}$ cm$^{-2}$ keV$^{-1}$
and $\Gamma$ is the {\it photon index}. However, X-ray spectra of
blazars show some deviations from the simple power-law. These
spectral signatures appear in the soft X-ray band with a curvature
(a flattening or a steepening of the spectrum); in addition, the
presence of Compton reflection components in RL objects is still
debated.

A clear physical interpretation of these features has not yet been
found. In the present work, we investigate these still open
questions in order to achieve a better understanding of the
general properties of the blazars spectra. The state-of-art of the
study of these important issues can be summarize as follows:


\begin{enumerate}

\item{Two physical interpretations to the flattening of the
primary intrinsic continuum with respect to a simple power-law
exist so far \citep{b37,b17,b54}: absorption in excess of the
Galactic component \citep[e.g.][]{b8,b54} and a break in the
intrinsic continuum \citep{b43,b46}. For the absorption excess, a
dependence between the hydrogen column density of the absorber
($N_H$) and the redshift $z$ has been proposed. In fact in a
former work by \citet{b54} based on XMM observations of a sample
of 32 RL sources, a $N_H$-$z$ correlation has been found.}

\item{Contrary to Radio Quiet (RQ) sources the presence of
reflection features - most notably the Fe $K\alpha$ line and the
associated Compton reflection "hump" at about $20-30$ keV - from
the cores of the RL AGN is not well established. In particular,
the reprocessed features are generally intense and always detected
in RQ AGN \citep{b34}, whereas in RL AGN the iron line and the
reflection component can be absent or weak.
X-ray observations of RL AGN with ASCA, RXTE and Beppo SAX
\citep[e.g.][]{b40,b11,b21,b3,b22,b23} have shown that some of
these objects (defined as Broad Line Radio Galaxies) seem to have
weak hard X-ray reflection features.

}

\end{enumerate}

In this paper we present a broad-band ($0.2-100$ keV) study of a
sample of 9 blazars up to redshift $\sim4$ observed with INTEGRAL,
Swift and XMM-Newton satellites. The wide energy range covered by
the instruments is well suited to better address their spectral
behaviour.

In Sections 2.1-2.4 we present the data set and analysis, while in
Section 3.1 we show the details of the spectral analysis. The
results are discussed in Section 3.2 and in Section 3.3 a study of
the distribution of different spectral parameters and possible
correlation is discussed and finally, we summarize the results in
Section 4.

\section[]{OBSERVATIONS AND DATA REDUCTION}

\subsection{THE HARD-X RAYS SELECTED SAMPLE}

The INTEGRAL/IBIS total sample is derived from the third
IBIS/ISGRI survey catalog that included 421 sources for a total
exposure time of 40Ms; 131 sources out of 421 have been identified
as AGN. The survey input data set consists of all pointing data
available at the end of 2006 May, from revolutions 12-429
inclusive, covering the time period from launch (17 October 2002)
to the end of 2006 April. Details about this survey are presented
by Bird et al. (2007). A subset of sources has been selected
according with the following selection criteria:

\begin{itemize}

\item{sources classified as blazars, optically identified
\citep[e.g.][]{b33,b64}}

\item{sources also observed in the soft X-ray energy domain
($0.2-10$ keV);}


\end{itemize}

Two further sources 3C 273 and 4C 04.42 have been excluded since
widely discussed elsewhere \citep{b56,b57}.

Our sample of INTEGRAL selected blazars, is composed of 9 sources:
7 FSRQ objects with $0.53 < z < 3.67$ and 2 BL Lac objects with
$z\sim0.07$ and $0.09$, but only five out of seven FSRQ were
available at the time of our analysis. For the 2 other objects we
adopted the hard X-rays BAT data.

Relevant information on the sample are reported in Table 1 where
we list the source name, type, optical coordinates, redshift,
Galactic absorption along the line of sight according to Dickey \&
Lockman (1990). In the last column of Table 1 we also list the
IBIS $20-100$ keV flux as reported by Bird et al. (2007) and the
BAT $14-195$ keV flux as reported by AGN Catalog of the first 9
months available at the time we started this project
\footnote{ The BAT data have been taken from the on-line archive
at: http://swift.gsfc.nasa.gov/docs/swift/results/bs9mon. For the
BAT spectra (related to PKS 2149-306 and 0537) we stress that in
the 22-months survey now available, the 15-45 keV fluxes are
doubled with respect to the 9-months survey. This effect is taken
into account with the cross calibration constant that has been
added when dealing with the fit between soft-X and hard-X rays
spectra. }.

We remark that source with IGR name (IGR J22517+2218) is
discovered in hard X-ray with INTEGRAL. It is the farthest object
so far detected by INTEGRAL whose nature was determined {\it a
posteriori } through optical spectroscopy \citep{b4}. It is worth
to note that two sources (FSRQs 3C 279 and PKS 1830-211) have been
detected in the Gamma Ray band with the Astro-rivelatore Gamma a
Immagini Leggero (AGILE) \footnote{ An Italian Space Agency (ASI)
mission launched on 23 April 2007 with a key scientific project:
the Gamma-Ray observations of blazars.} \citep{b19,b45,b62} and
the Large Area Telescope on board the Fermi Gamma-ray Large Area
Space Telescope (GLAST) \footnote{An international and
multi-agency mission launched on 11 June 2008; one of its major
scientific goals is to provide new data on the Gamma-ray activity
of AGN.}. The latter satellite also revealed a third source
belonging to our sample, the object BL Lac \citep{b1}.



\begin{table*}
\begin{center}
\begin{flushleft}

\tiny

\caption{Data for our sample of blazars observed with INTEGRAL,
Swift and XMM-Newton.}
\begin{tabular}{lccccccccccc}
\noalign{\hrule} \noalign{\medskip}

\multicolumn{8}{l}{}\\
 $Source$ & $Type$ & $Broad-Band$ & R.A. & Dec & $redshift$ & $N_H^{Gal}$ & $F_{20-100keV}^{IBIS}$ & $L_{20-100keV}^{rest-frame}$ \\
\\
 &  &  &  &  &  & [$10^{22}cm^{-2}$] & ($F_{14-195keV}^{BAT}$) & ($L_{14-195keV}^{rest-frame}$)\\
 \\

 &  &  &  & & &  & [$10^{-11}erg\cdot cm^{-2}\cdot s^{-1}$] & [$10^{46}erg\cdot s^{-1}$] \\
\\
\hline
\\

 QSO B0836+710 & FSRQ & XMM+INTEGRAL & 08 41 24.37 & +70 53 42.2 & 2.172 & 0.030 &  6  & 63.1 \\
\\

 1ES 0033+595 & BL Lac & XMM+INTEGRAL & 00 35 52.63 & +59 50 04.6 & 0.086 & 0.427 &  2 & 0.1  \\
\\

 PKS 0537-286 & FSRQ & XMM+BAT & 05 39 54.28 & -28 39 55.9 & 3.104 & 0.020 &  (3) &  (31.6) \\
\\

 PKS 2149-307 & FSRQ & XMM+BAT & 21 51 55.52 & -30 27 53.7 & 2.345 & 0.020 &  (5) &  (39.8) \\
\\

 Swift J1656.3-3302 & FSRQ & XRT+INTEGRAL & 16 56 16.56 & -33 02 09.3 & 2.4 & 0.220 &  2 &  50.1 \\
\\

 IGR J22517+2218 & FSRQ & XRT+INTEGRAL & 22 51 53.50 & +22 17 37.3 & 3.668 & 0.050 &  4 &  100.0 \\
\\

 Bl Lac & BL Lac & XRT+INTEGRAL & 22 02 42.72 & +42 17 16.8 & 0.069 & 0.022 &  3 &  0.1 \\
\\

 PKS 1830-211 & FSRQ & XMM+INTEGRAL & 18 33 39.89 & -21 03 39.8 & 2.507 & 0.260 &  5 &  100.0 \\

\\

 3C 279 & FSRQ & XRT+INTEGRAL & 12 56 11.17 & +05 47 21.5 & 0.536 & 0.020 & 2 & 2.5 \\
\\

\hline

\end{tabular}
\label{Silvia1}
\end{flushleft}

\end{center}

\end{table*}


\subsection{INTEGRAL DATA}

The INTEGRAL data presented here are based on pointings with the
IBIS instrument \citep{b47}, collected over the period from end of
2002 up to April 2006 (revolution 12 up to 429). Images from the
ISGRI detector \citep{b29} for each pointing have been genereated
in different energy bands using off-line scientific Analysis
Software \citep{b20} OSA version 5.1. Count rates at the source
position have been extracted from individual images to provide
light curves in different energy bands. From light curves the
average fluxes have been derived and combined to produce an
average source spectrum \citep[for details see][]{b6} in the
$20-100$ keV band.

\subsection{XMM-NEWTON DATA}

XMM data of the six blazars (1ES 0033+595, 4C 04.42, PKS 1830-211,
QSO B0836-710, PKS 0537-286 and PKS 2149-307) are a combination of
proprietary data and public observations obtained from the
XMM-Newton Science
Archive\footnote{http://xmm.esac.esa.int/xsa/index.shtml.}. The
raw EPIC Observation Data Files (ODFs) were obtained from the XMM
Science Archive and reduced using the standard Science Analysis
System (SAS) software package (v.7.1.0) and the most recent
calibration files available at the time of the data reduction. We
used the EMCHAIN and EPCHAIN task for the pipeline processing of
the ODFs to generate the corresponding event files. The spectra
were created using X-ray events of pattern 0-12 for MOS and 0-4
for PN.
The source counts were extracted from a circular region centred on the
source with a radius of 20-40 arcsec and the background was
derived from two nearby source-free circular regions of the same
size. Spectra were re-binned using GRPPHA to have a minimum of 20
counts in each bin \footnote{10 counts are the minimum number of
counts required to use the $\chi^2$ minimization technique.}, so
that the $\chi^2$ statistic could reliably be used. Details of the
XMM-Newton observations and the source observed count rates are
summarized in Table 2.



\begin{table*}
\begin{center}
\begin{flushleft}
\caption{XMM-Newton observations log and source observed count
rates.}
\begin{tabular}{lcccccccc}
\noalign{\hrule} \noalign{\medskip}

\multicolumn{9}{l}{}\\
 $Source$ & Obs.ID & Date of obs. & Exp. time & Exp. time & Exp. time & Counts/s & Counts/s & Counts/s \\

 &  &  & MOS1 & MOS2 & PN & MOS1 & MOS2 & PN \\
 &  &  &  &  &  & ($0.2-10 keV$) & ($0.2-10 keV$) & ($0.2-10 keV$) \\
 &  &  & $(s)$ & $(s)$ & $(s)$ & $(s^{-1})$ & $(s^{-1})$ & $(s^{-1})$ \\

\\
\hline
\\

 1ES 0033+595 & 0094381301 & 2003-02-01 & 1204 & 1535 & 4209 & $ 1.39\pm0.04 $ & $ 1.39\pm0.04 $ & $ 3.61\pm0.03 $ \\

\\

 PKS 1830-211 & 0204580301 & 2004-03-24 & 30445 & 30444 & 20906 & $ 0.58\pm0.01 $ & $ 0.58\pm0.01 $ & $ 1.65\pm0.01 $ \\

\\

 QSO B0836-710 & 0112620101 & 2001-04-12 & 35614 & 35620 & 29035 & $ 3.88\pm0.01 $ & $ 3.85\pm0.01 $ & $ 11.97\pm0.02 $ \\

\\

 PKS 0537-286 & 0206350101 & 2005-03-20 & 80610 & 80765 & 64661 & $ 0.205\pm0.002 $ & $ 0.226\pm0.002 $ & $ 0.746\pm0.004 $ \\

\\

 PKS 2149-307 & 0103060401 & 2001-05-01 & 23959 & 23960 & 20287 & $ 0.63\pm0.01 $ & $ 0.62\pm0.01 $ & $ 1.67\pm0.01 $ \\

\\

\hline

\end{tabular}
\label{Silvia1}
\end{flushleft}

\end{center}

\end{table*}


\subsection{SWIFT/XRT DATA}

Swift/XRT data reduction of four blazars (3C 279, BL Lac, IGR
J22517+2218, Swift J1656.3-3302) was performed using the tool
XSELECT v. 2.4. Events for spectral analysis were extracted within
a circular region of radius 20 arcsec centred on the source
position. The background was extracted from a circular region with
the same radius and located far off the source. In all cases, the
spectra were binned using GRPPHA. We used Ancillary Response Files
(ARFs) and Response Matrix Files (RMFs) available for download
from the HEASARC Calibration Database (caldb) calibration files
at:
$http://heasarc.gsfc.nasa.gov/docs/heasarc/caldb/caldb\_intro.html$.
Table 3 lists the observation ID, the observation start time, the
integration time and the observed count rate.



\begin{table*}
\begin{center}
\begin{flushleft}
\caption{Swift/XRT observations log and source observed count
rates.}
\begin{tabular}{lcccc}
\noalign{\hrule} \noalign{\medskip}

\multicolumn{5}{l}{}\\
$Source$ & Obs.ID & Date of obs. & Exp. time & Counts/s \\

 &  &  & XRT & XRT \\
 &  &  &  & ($0.2-10 keV$) \\
  &  &  & $(s)$ & $(s^{-1})$ \\
\\
\hline
\\
 BL Lac & 00090042010 & 2008-08-29 & 5806 & $ 0.20\pm0.01 $ \\
\\

 IGR J22517+2218 & 00036660001 & 2007-05-26 & 8300 & $ 0.054\pm0.003 $ \\
\\

 Swift J1656.3-3302 & 00035272002 & 2006-06-13 & 4785 & $ 0.075\pm0.004 $ \\
\\

 3C 279 & 00035019009 & 2008-11-26 & 22580 & $ 0.198\pm0.003 $ \\
\\

\hline

\end{tabular}
\label{Silvia1}
\end{flushleft}

\end{center}

\end{table*}


%


\begin{figure*}
\begin{minipage}{1\linewidth}
\centering
\includegraphics[height=14cm,width=1\linewidth]{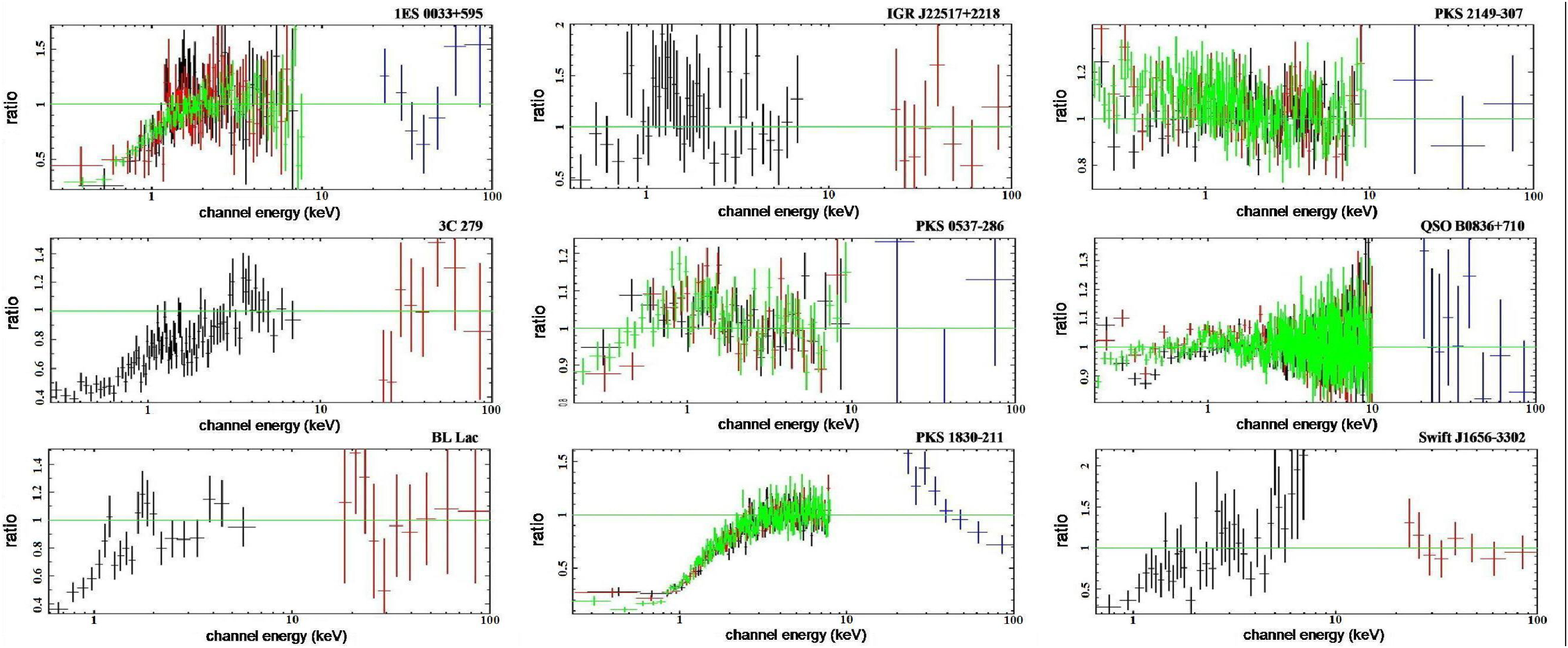}
\caption{The data to model $A$ ratio as obtained by fitting a simple power-law over the observed frame $3-100$ keV energy range, and extrapolating over the $0.2-100$ keV range. The black, red and green bins up to 10 keV are the data of MOS1, MOS2 and PN instruments, respectively.}
\end{minipage}
\end{figure*}

\section{BROAD-BAND SPECTRAL ANALYSIS}

The spectral analysis was performed by combining the different
data from each instrument using the X-ray SPECtral fitting
software XSPEC v. 11.3.2 \citep{b2}. The errors are quoted at 90\%
confidence level for one interesting parameter. The energy values
are always given in the observer frame, if not otherwise
specified. Throughout this paper, a WMAP (Wilkinson Microwave
Anisotropy Probe) Cosmology of $H_0=70$ $km$ $s^{-1}$ $Mpc^{-1}$,
$\Omega_{\lambda}=0.73$ and $\Omega_m=1-\Omega_{\lambda}$ is
assumed.

We have combined XMM spectrum with the IBIS one for the blazars
1ES 0033+595, 4C 04.42, PKS 1830-211, QSO B0836-710; XRT spectrum
with the IBIS one for 3C 279, BL Lac, IGR J22517+2218, Swift
J1656.3-3302 and XMM spectrum with the BAT one for PKS 0537-286,
PKS 2149-307 (see Table 1).

To account for possible cross-calibration mismatches between
different instruments as well as to take into account flux
variations between the observing periods, a multiplicative
constant factor (C) has always been added to the fit. In general,
the calibration between the various satellites is performed
analyzing the Crab spectrum. In particular, various studies of the
Crab spectrum \citep[e.g.][]{b26,b27} have shown that the
inter-calibration XMM/INTEGRAL and Swift/INTEGRAL is close to 1
(within a few percent). We remark that, since XMM, Swift and
INTEGRAL observations were not simultaneous, flux variations are
also possible according to the blazar-nature of objects. Being the
inter-calibration constant C $\sim1$, we suggest that, a different
value for this constant is likely due to flux variability. In our
analysis, this constant C between the two instruments at low and
high energies has been allowed to vary freely in the fits. In
particular, we have considered the MOS1 and XRT instruments as
reference detectors in the case of XMM-INTEGRAL or BAT data and
XRT-INTEGRAL data, respectively. Therefore, the inter-calibration
constants in our fitting procedure are MOS2/MOS1, PN/MOS1,
INTEGRAL/MOS1, BAT/MOS1 and INTEGRAL/XRT. However MOS2/MOS1 and
PN/MOS1 are always $\sim1$ within few percent.

\subsection{SPECTRAL FITTING RESULTS}

To reproduce the shape of the intrinsic continuum of our sources
we fitted the combined spectra ($0.2-100$ keV) of each source with
several models.

Firstly, we fit the spectra of our sources in the $3-100$ keV
energy band with the model A, that corresponds to a simple
power-law absorbed by gas of the Milky Way. The results of the fit
are summarized in Table 4. The model A provides a good fit for all
sources. When extrapolating this model to the low energy range, a
very poor fit in all cases but one (PKS 2149-307) was obtained; in
fact the data to model ratio (see Figure 1) shows a clear
systematic residuals below $\sim3$ keV in the observer frame. From
the ratios shown in Figure 1, we can see that the spectra of the
majority of the sources clearly show a deficit of soft X-ray
counts with respect to the simple power-law model, indicating the
possible presence of excess absorption and/or of an energy break
in the intrinsic continuum.

In order to properly reproduce such a feature we tested the impact
of different models.

In the model B we added an extra-absorption in the rest-frames of
the sources (phabs*zphabs*zpo model in XSPEC) and results are
presented in Table 5.

All sources of our sample, with the exception of PKS 2149-307 and
PKS 1830-211, were well fitted with the model B, while for the
source PKS 2149-307 we have obtained an upper limit value on the
intrinsic absorbing column density and for PKS 1830-211 only a
poor quality fit to the data has been derived. Details on this
source will be discussed later in this section.

Then, we tested a broken power-law continuum model C (phabs*bknpo
model in XSPEC) for the whole sample. Results are shown in the
Table 6. This model did not improve the fit further in most of the
cases and, also in those cases for which a slightly better
$\chi^2$ is obtained, the null hypothesis probability increases
only a few per cent with respect to model B. Although strong
statistical evidence does not exist for a broken power-law respect
to the absorbed one, however the presence of an intrinsic spectral
break cannot be ruled out.

For PKS 1830-211, both models B and C fail to reproduce source
data. PKS 1830-211 is a gravitationally lensed galaxy and the
excess absorption can be attributed to the intervening galaxy at
redshift $z=0.89$. The best-fit of the observed spectrum requires
both an absorber and a broken power-law (model D), confirming the
results reported by \citet{b51} with three XMM-Newton observations
Obs.ID 0204580201, 0204580301 and 0204580401\footnote{In our
analysis we have chosen the observation with longer exposure
time.} and the INTEGRAL average data (up to April 29, 2006). In
fact, in addition to the excess cold absorption, PKS 1830-211 also
exhibits an energy break at about $4$ keV in the observer frame.
Moreover, our result for the absorption in excess to the Galactic
one, is in agreement with that obtained previously by \citet{b63}
with Chandra observations and a subset of the actual IBIS data.
Table 7 reports the results of our best-fit model.




Finally, we also checked a Compton reflection component with a
power-law reflected from neutral matter (pexrav model in XSPEC,
\citet{b31}); the reflection component is parameterized in terms
of $R$=$\Omega/{2\pi}$, the solid angle in units of $2\pi$
subtended by the reflecting matter, assumed to be observed
face-on). This model resulted in a worst fit and upper limit
values on $R$ for all absorbed sources (see Table 8).

In addition, we have searched for the presence of an iron line
emission in the spectra of our sources. To this aim, we adopted
the following method: we exclude the Fe K band, fit a continuum,
then freeze the continuum, after we include the Fe K band and
perform the fit again. The comparison between the values of the
$\chi^2$ of the fits performed including and excluding the Fe K
band provides us a clear indication for the real presence of this
feature of the observed spectra.

For all sources, but one (IGR J22517+2218), the application of
this method does not produce any clear evidence for the presence
of this feature in the spectrum. The main results for the source
IGR J22517+2218 can be summarized as follows: the first step
described before resulted in a photon index $\Gamma$=1.5 and an
upper limit on $N_{H}$=5$\times$10$^{22}$ cm$^{-2}$ with
${\chi}_r^2$/dof=23/29; in the second step we included the iron
channels and then we added a narrow line (${\sigma}$=0.01 keV) and
the fit has been repeated obtaining ${\chi}_r^2$/dof=37/44 with a
probability $P_{Ftest}>99.7\%$ implying that this feature is real
with an Equivalent Width EW=$80\pm50$ eV; repeating the fit with a
free photon index a value of $\Gamma=1.6\pm0.2$ is obtained with
an iron line that is significant at the 99\%. In summary, for this
source there is the evidence for the presence of a weak line
significant at the 99\% level.

As a conclusion, we firmly asses that the an iron emission line is
not request by our data, with the exception of the source IGR
J22517+2218.





\begin{table*}
\begin{center}
\begin{flushleft}
\caption{Best fit parameters of a simple power-law model in 3-100
keV energy range.}
\begin{tabular}{lcccccccccc}
\noalign{\hrule} \noalign{\medskip}

\multicolumn{7}{l}{Model $A$}\\

\\

Name & $^1$$F_{ob(3-10keV)}$ & $^1$$F_{ob(20-100keV)}$ & $^3$$\Gamma$ & $^{4}$C & $^5$${\chi}_r^2$/dof & $^6$$P_{null}$ \\
\hline
\\

 QSO B0836+710 & 4.1 & 5.2 & $1.34\pm0.02$ & $0.28\pm0.03$ & 0.91/1521 & 0.996 \\
\\

 1ES 0033+595 & 1.1 & 1.5 & $2.52^{+0.18}_{-0.17}$ & $4.2^{+2.5}_{-1.6}$ & 0.79/53 & 0.865 \\
\\

PKS 0537-286 & 0.5 & 2.3 & $1.19\pm0.06$ & $0.6\pm0.2$ & 1.05/329 & 0.253 \\
\\

PKS 2149-307 & 0.6 & 4.9 & $1.41\pm0.06$ & $1.9\pm0.5$ & 0.91/225 & 0.835 \\
\\

PKS 1830-211 & 1.4 & 4.4 & $1.22\pm0.04$ & $0.5\pm0.1$ & 1.03/442 & 0.315 \\
\\

Swift J1656.3-3302 & 0.5 & 2.1 & $1.60^{+0.33}_{-0.32}$ & $1.6^{+2.0}_{-0.9}$ & 1.02/18 & 0.434 \\
\\

IGR J22517+2218 & 0.3 & 3.8 & $1.43^{+0.59}_{-0.56}$ & $4.1^{+11.4}_{-3.1}$ & 0.66/14 & 0.817 \\
\\

Bl Lac & 0.6 & 2.3 & $2.01^{+0.43}_{-0.41}$ & $4.3^{+6.5}_{-2.6}$ & 0.59/11 & 0.840 \\
\\

3C 279 & 0.5 & 2.1 & $2.00^{+0.24}_{-0.23}$ & $4.3^{+3.1}_{-1.8}$ & 1.06/19 & 0.382 \\
\\

\hline

\end{tabular}
\label{Silvia1}
\end{flushleft}

\end{center}

\small{

$^1$ Observed flux in the $2-10$ keV (or $20-100$ keV) energy
range in units of $\rm 10^{-11}$ erg cm$^{-2}$ s$^{-1}$. $^3$
Photon index, related to the spectral index $\alpha$ (where
$F_\nu\propto{\nu^{-\alpha}}$) by $\alpha=\Gamma-1$. $^4$
IBIS/MOS, BAT/MOS or IBIS/XRT cross-calibration constant. $^5$
Reduced chi-squared to degrees of freedom. $^6$ Null hypothesis
probability that is the probability of getting a value of $\chi^2$
as large or larger than observed if the model is correct. If this
probability is small then the model is not a good fit to the data.
}

\end{table*}


\begin{table*}
\begin{center}
\begin{flushleft}
\caption{Best fit parameters of an absorbed power-law model in
0.2-100 keV energy range.}
\begin{tabular}{lccccccccccc}
\noalign{\hrule} \noalign{\medskip}

\multicolumn{8}{l}{Model $B$}\\

\\

Name & $^1$$F_{ob(2-10keV)}$ & $^1$$F_{ob(20-100keV)}$ & $^2$$N_{H}$ & $^3$$\Gamma$ & $^{4}$C & $^5$${\chi}_r^2$/dof & $^6$$P_{null}$ \\
\hline
\\

 QSO B0836+710 & 4.0 & 4.9 & $0.07\pm0.02$ & $1.353\pm0.005$ & $0.28\pm0.03$ & 1.02/2421 & 0.206 \\
\\

 1ES 0033+595 & 1.2 & 1.5 & $0.20\pm0.02$ & $2.47\pm0.05$ & $3.5\pm0.8$ & 0.81/229 & 0.985 \\
\\

PKS 0537-286 & 0.5 & 2.3 & $0.30\pm0.05$ & $1.22\pm0.01$ & $0.7\pm0.2$ & 1.03/972 & 0.235 \\
\\

PKS 2149-307 & 0.6 & 4.8 & $<0.05$ & $1.45\pm0.01$ & $2.1\pm0.5$ & 0.99/849 & 0.543 \\
\\

Swift J1656.3-3302 & 0.5 & 2.0 & $7.9^{+5.1}_{-3.8}$ & $1.66^{+0.22}_{-0.20}$ & $1.9^{+1.7}_{-0.9}$ & 1.03/42 & 0.422 \\
\\

IGR J22517+2218 & 0.2 & 3.5 & $3.6^{+3.3}_{-2.2}$ & $1.66^{+0.21}_{-0.19}$ & $7.1^{+6.7}_{-3.4}$ & 0.85/49 & 0.770 \\
\\

Bl Lac & 0.5 & 2.3 & $0.20\pm0.07$ & $2.01^{+0.17}_{-0.16}$ & $4.3^{+2.8}_{-1.7}$ & 0.81/31 & 0.762 \\
\\

3C 279 & 0.5 & 2.2 & $0.04\pm0.02$ & $1.75\pm0.06$ & $2.5^{+0.7}_{-0.6}$ & 1.07/79 & 0.311 \\
\\

\hline

\end{tabular}
\label{Silvia1}
\end{flushleft}

\end{center}

\small{ $^1$ Observed flux in the $2-10$ keV (or $20-100$ keV)
energy range in units of $\rm 10^{-11}$ erg cm$^{-2}$ s$^{-1}$.
$^2$ Intrinsic column density of hydrogen in units of $\rm
10^{22}$ cm$^{-2}$. $^3$ Photon index, related to the spectral
index $\alpha$ (where $F_\nu\propto{\nu^{-\alpha}}$) by
$\alpha=\Gamma-1$. $^4$ IBIS/MOS, BAT/MOS or IBIS/XRT
cross-calibration constant. $^5$ Reduced chi-squared to degrees of
freedom. $^6$ Null hypothesis probability that is the probability
of getting a value of $\chi^2$ as large or larger than observed if
the model is correct. If this probability is small then the model
is not a good fit to the data. }

\end{table*}


\begin{table*}
\begin{center}
\begin{flushleft}
\caption{Best fit parameters of a broken power-law model in
0.2-100 keV energy range.}
\begin{tabular}{lccccccccccc}
\noalign{\hrule} \noalign{\medskip}

\multicolumn{8}{l}{Model $C$}\\

\\

Name & $^1$$F_{ob(2-10keV)}$ & $^1$$F_{ob(20-100keV)}$ & $^7$$\Gamma_{1}$,$\Gamma_{2}$ & $^{8}$$E_{b}$ & $^{4}$C & $^5$${\chi}_r^2$/dof & $^6$$P_{null}$ \\
\hline
\\

 QSO B0836+710 & 4.0 & 4.9 & $1.26^{+0.01}_{-0.02}$ & $0.98\pm0.08$ & $0.28\pm0.03$ & 1.02/2420 & 0.274 \\

   &  &  & $1.354\pm0.005$ &  &  &  &  \\
\\

 1ES 0033+595 & 1.2 & 1.6 & $1.38^{+0.16}_{-0.13}$ & $1.39^{+0.12}_{-0.10}$ & $2.8^{+0.7}_{-0.6}$ & 0.85/228 & 0.947 \\

   &  &  & $2.37^{+0.06}_{-0.05}$ &  &  &  &  \\
\\

PKS 0537-286 & 0.5 & 2.3 & $0.83^{+0.09}_{-0.12}$ & $0.72\pm0.09$ & $0.7\pm0.2$ & 1.03/971 & 0.282 \\

   &  &  & $1.21\pm0.01$ &  &  &  &  \\
\\

PKS 2149-307 & 0.6 & 4.8 & $1.47^{+0.01}_{-0.02}$ & $2.7^{+4.5}_{-1.1}$ & $1.9\pm0.5$ & 0.99/848 & 0.579 \\

   &  &  & $1.41^{+0.04}_{-0.34}$ &  &  &  &  \\
\\

Swift J1656.3-3302 & 0.6 & 2.0 & $1.03^{+0.17}_{-0.47}$ & $8.3^{+45.5}_{-3.4}$ & $1.8^{+2.2}_{-1.4}$ & 0.87/41 & 0.708 \\

   &  &  & $1.96^{+0.30}_{-0.27}$ &  &  &  &  \\
\\

IGR J22517+2218 & 0.2 & 3.5 & $0.58^{+0.49}_{-0.69}$ & $1.19^{+0.27}_{-0.26}$ & $7.8^{+8.4}_{-3.7}$ & 0.82/48 & 0.803 \\

   &  &  & $1.69^{+0.23}_{-0.19}$ &  &  &  &  \\
\\

Bl Lac & 0.5 & 2.2 & $1.06^{+0.57}_{-0.73}$ & $1.71^{+0.75}_{-0.15}$ & $4.6^{+1.7}_{-1.0}$ & 0.89/30 & 0.633 \\

   &  &  & $2.02^{+0.34}_{-0.12}$ &  &  &  &  \\
\\

3C 279 & 0.5 & 2.2 & $1.46^{+0.10}_{-0.11}$ & $1.16^{+0.27}_{-0.20}$ & $2.7^{+0.8}_{-0.7}$ & 1.01/78 & 0.446 \\

   &  &  & $1.78\pm0.06$ &  &  &  &  \\
\\

\hline

\end{tabular}
\label{Silvia1}
\end{flushleft}

\end{center}

\small{ $^1$ Observed flux in the 2-10 keV (or 20-100 keV) energy
range in units of $\rm 10^{-11}erg cm^{-2} s^{-1}$. $^4$ IBIS/MOS,
BAT/MOS or IBIS/XRT cross-calibration constant. $^5$ Reduced
chi-squared to degrees of freedom. $^6$ Null hypothesis
probability that is the probability of getting a value of $\chi^2$
as large or larger than observed if the model is correct. If this
probability is small then the model is not a good fit to the data.
$^7$ Spectral index below the break ($\Gamma_{1}$) and above the
break ($\Gamma_{2}$). $^{8}$ Observed break energy in units of
keV.
 }

\end{table*}


\begin{table*}
\begin{center}
\begin{flushleft}
\caption{Best fit parameters of a broken power-law model with an
absorption in excess in 0.2-100 keV energy range.}
\begin{tabular}{lccccccccccc}
\noalign{\hrule} \noalign{\medskip}

\multicolumn{9}{l}{Model $D$}\\

\\

Name & $^1$$F_{ob(2-10keV)}$ & $^1$$F_{ob(20-100keV)}$ & $^2$$N_{H}$ & $^7$$\Gamma_{1}$,$\Gamma_{2}$ & $^{8}$$E_{b}$ & $^4$C & $^5$${\chi}_r^2$/dof & $^6$$P_{null}$ \\

\hline
\\
 PKS 1830-211 & 1.4 & 4.3 & $1.5^{+0.2}_{-0.1}$ & $0.95\pm0.05$ & $4.01^{+0.42}_{-0.54}$ & $0.8^{+0.3}_{-0.2}$ & 1.06/856 & 0.101 \\

 &  &  &  & $1.32^{+0.10}_{-0.07}$ &  &  &  &  \\

\hline

\end{tabular}
\label{Silvia1}
\end{flushleft}

\end{center}

\small{ $^1$ Observed flux in the $2-10$ keV (or $20-100$ keV)
energy range in units of $\rm 10^{-11}$ erg cm$^{-2}$ s$^{-1}$.
$^2$ Intrinsic column density of hydrogen in units of $\rm
10^{22}$cm$^{-2}$. $^4$ IBIS/MOS, BAT/MOS or IBIS/XRT
cross-calibration constant. $^5$ Reduced chi-squared to degrees of
freedom. $^6$ Null hypothesis probability that is the probability
of getting a value of $\chi^2$ as large or larger than observed if
the model is correct. If this probability is small then the model
is not a good fit to the data. $^7$ Spectral index below the break
($\Gamma_{1}$) and above the break ($\Gamma_{2}$). $^{8}$ Observed
break energy in units of keV. }

\end{table*}



\begin{table*}
\begin{center}
\begin{flushleft}

\tiny

\caption{Upper limits to $R$ under the assumption of a pexrav
model.}
\begin{tabular}{lccccccccc}
\noalign{\hrule} \noalign{\medskip}


Name & QSO B0836+710 & 1ES 0033+595 & PKS 0537-286 & PKS 2149-307 & PKS 1830-211 & Swift J1656.3-3302 & IGR J22517+2218 & Bl Lac & 3C 279 \\
\hline

\\

 $R$ & 0 & 0 & 0.09 & 0.08 & 0 & 0.5 & 3E-6 & 0 & 3E-4 \\

\\

\hline

\end{tabular}
\label{Silvia1}
\end{flushleft}

\end{center}

\end{table*}


\subsection{DISCUSSION}

Our broad-band spectral analysis has shown that a deficit of
photons in the soft X-ray energy domain is a common feature in our
sample. This phenomenon has been previously observed in several
samples of RL objects \citep[e.g.][]{b37,b17,b54} and its origin
is not yet well understood. It can be produced by either an
intrinsically curved continuum or an extra absorption component.

The broken continuum hypothesis is linked to the blazar-type
source. In fact, in the hypothesis of an intrinsically curved
blazar continuum, one can assume that the relativistic electrons
in the jet follow a broken power-law energy distribution with a
low energy cut-off, producing a flattening in the observed
spectrum. This is a purely phenomenological form assumed to
reproduce the observed shape of the blazar SEDs, without any
specific assumption on the acceleration/cooling mechanism acting
on the particles \citep[e.g.][]{b46}. Alternatively, it can be
also assumed there is an inefficient radiative cooling of lower
energy electrons of the jet producing a few Synchrotron photons to
be scattered at higher energies \citep[e.g.][]{b43}.

We emphasize that on basis of our spectral analysis an intrinsic
spectral break cannot be ruled out (see the values of the reduced
$\chi^2$ and of the null hypothesis probability listed in the
Tables 5 and 6). In this context, we just noted that the existence
of a break in the intrinsic continuum is tightly related to the
electron distribution, but
important constraints on the curved electron distribution can be
derived only analyzing both Synchrotron and EC components in the
SED, that is out of the aim of present work. Therefore we will
focus the following discussion on the absorption scenario.


The absorption explanation takes into account an absorbing gas in
excess of the Galactic one; in some cases, the absorbing medium
can be cold (not ionized) and located either between the observer
and the source along the line of sight at a different redshift
with respect to the blazar \citep[e.g.][]{b8,b16,b13,b33}, or at
the same redshift of the source \citep[e.g.][]{b37,b54} and it is
namely an intrinsic cold absorber.

For an absorber located at several redshifts between the observer
and the source, the damped Ly-alpha systems are the more plausible
cause \citep[e.g.][]{b10} and the line of sight would have to pass
through two or more very high column density damped Ly-alpha
systems \citep[e.g.][]{b14}. However, this is a rare occurrence
\citep[e.g.][]{b35,b52} and favours the hypothesis of an intrinsic
absorber with column densities of the order $10^{22}-10^{23}$
cm$^{-2}$. An intrinsic origin is also confirmed by the soft X-ray
spectral flattening detected so far only in radio loud objects
\citep{b5,b38,b42}.

The intrinsic hypothesis also suggests that the medium could be
"warm" at the same stage. In this case, the absorbing medium, at
the same redshift of the AGN (probably, the inner part of the
dense interstellar medium of the young host galaxy), is ionized as
due to the proximity from the jet \citep[e.g.][]{b13,b54}.

For our sample, the fitting procedure supports the presence of an
intrinsic (local to the AGN) cold absorber for all sources.
Nevertheless a warm absorber model (absori in XSPEC) can be
excluded since the model with an ionized-gas produces a worse fit
and an inconsistent value for the ionization parameter ($\xi\sim0$
$erg$ $cm$ $s^{-1}$ and $\xi$ $> 500$ $erg$ $cm$ $s^{-1}$, values
indicating a neutral absorber and a completely ionized absorber,
respectively).

The information on the absorption is of crucial importance to
understand the blazar environment and its interaction with the
jet.

Absorption originating from the material present in the AGN
environment
seems to be a more convincing explanation for many sources
belonging to the class of RL AGN \citep[][and references
therein]{b37,b54}. There is some evidence not only of absorption
but also of a correlation between absorption and redshift, with
the more distant sources being more absorbed and the increase in
$N_H$ with $z$ occurring starting at $z\sim2$ \citep{b54}. We note
that more than half of the sources in our sample appear in
agreement with this scenario, since this fraction of objects has a
redshift larger than 2.


As far it concerns the nature of this absorber, we suggest it
could be associated to the presence of gas in the host galaxy
illuminated by the blazar jet.


The excess of intrinsic absorption in the rest-frame of the RL
objects compared with the lack of absorption above the Galactic
value of their RQ counterparts
\citep{b7,b28,b53,b36,b37,b49,b50,b24}, suggests that an intrinsic
difference could exist in the local environments of the two
classes of objects. Since the X-ray spectra of RL AGN appear to be
dominated by the emission of the jet component, most likely the
absorbing material is linked to the presence of this component.
This working scenario is supported by a theory presented, e.g., in
\citet{b15}, in which the confinement of the radio jet needs the
presence of a cold medium (local to the AGN) so that jet expansion
is suppressed.

In our sample we have found neither an evidence of an iron
emission line at 6.4 keV nor evidence of a Compton reflection
"hump" above 10 keV.
The lack of these spectral features -- interpreted as the result
of reprocessing of the primary continuum by cold matter around the
X-ray source, presumably the accretion disk
\citep{b30,b25,b18,b39,b40} -- can be explained either with the
out-and-out absence of reprocessing features in blazar-type
objects or with the hypothesis that these features are difficult
to be detect in strong radio sources being likely dilute by the
jet \citep{b41}.

There is also evidence that the spectral shape does not change
between $0.2-10$ keV and $20-100$ keV energy ranges, supporting
the idea that, above $10$ keV, no additional component to the
power-law (e.g. Compton reflection hump) is present in the
spectrum.

The value of the cross-calibration constant (see Tables B and C)
shows that a strong flux variability is present in QSO B0836+710,
1ES 0033+595, PKS 1830-211, IGR J22517+2218, BL Lac, and 3C 279.
In particular, we derived that IGR J22517+2218 showed severe
intensity changes between the INTEGRAL and the Swift observations
with a value of the XRT/IBIS inter-calibration factor of about
$7$. This is not a surprise because of the nature of the object
and data being not simultaneous. In addition, we remark that
INTEGRAL/IBIS data set correspond to an average performed on
observations spanning on 3 years, while the data for the
low-energy range (XMM and XRT) correspond to observations with a
time baseline of few hours.


\subsection{PARAMETER DISTRIBUTIONS AND CORRELATIONS}

Although the number of sources of our sample is not large, we have
also analyzed the distribution of the spectral parameters
(absorbing column density, spectral index, redshift) and their
possible correlations. This type of analysis has been performed,
so far, only between $0.2-10$ keV \citep{b54}. Our study is still
preliminary and has been performed with a detailed comparison
among the results of our own spectral analysis and those already
available in literature in a limited energy range. The reference
sample includes 35 RL objects -- with 19/35 sources identified as
blazars -- \citep{b17,b37,b54} and shows an absorption of unknown
intrinsic nature. In order to avoid any problem of "orientation
dependence bias" when studying the $N_H$ and $\Gamma$
distribution, we have selected, from the quoted sample, only the
blazar-type objects. In particular, for the $N_H$ study we have
accounted for the blazars with a specific value of the intrinsic
absorption (we have not accounted for the upper limits on $N_H$),
obtaining a sub-sample of 16 objects.

In the framework of a possible $N_H$-redshift relation, we also
combined our results
 with the larger sample of RL objects
\citep{b17,b37,b54}, counting out three sources belonging to our
sample (QSO B0836+710 PKS 2149-307 and PKS 0537-286). In fact, it
is interesting to check whether the 9 selected blazars in the
sample and analyzed in a more wide energy band, follow the trend
of $N_H$ with the redshift previously found by \citet{b54}. If
this would be the case, we would have further evidence that the
absorbing column density, in these sources, is linked to the
distance and thus a confirmation of a cosmic evolution effect.

\subsubsection{ABSORPTION DISTRIBUTION}

We made a detailed analysis of the rest-frame hydrogen column
density for our selected blazars, in conjunction with the
intervening absorption in excess to the galactic one.

\begin{figure}
\includegraphics[width=64mm]{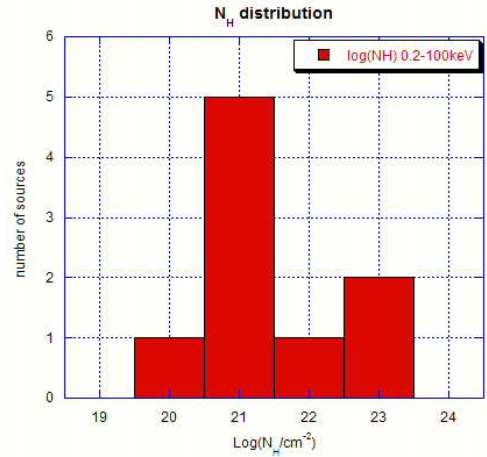}
 \caption{Distribution of the X-ray absorption column density $(N_H)$ for our sample objects over the $0.2-100$ keV energy range.}
\end{figure}

\begin{figure}
\includegraphics[width=64mm]{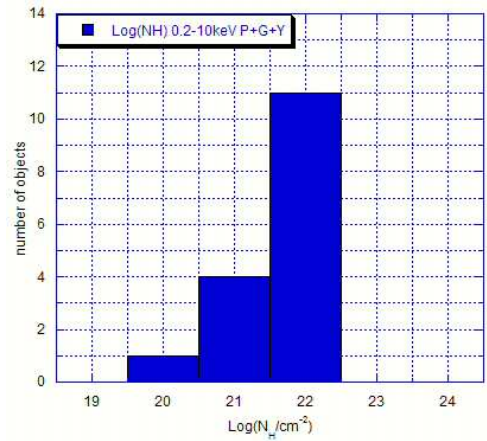}
 \caption{The $N_H$ distribution of a larger sample of 16 blazars over the $0.2-10$ keV energy range. See text for more details.}
\end{figure}



\begin{figure}
\includegraphics[width=85mm]{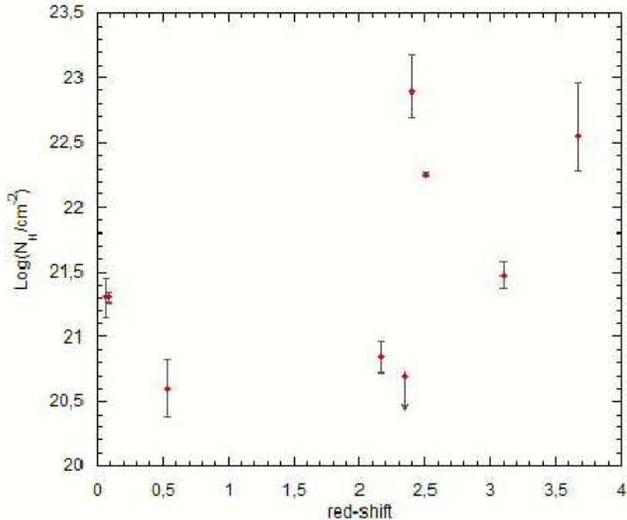}
 \caption{The X-ray absorption column density $N_H$ as a function of the redshift for our sample.}
\end{figure}

\begin{figure}
\includegraphics[width=85mm]{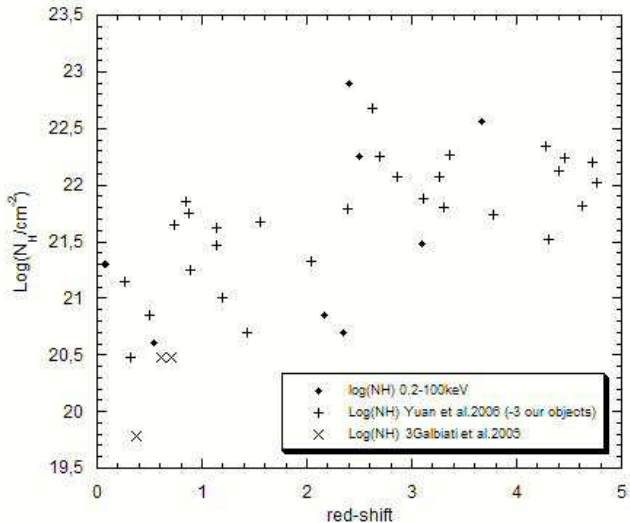}
 \caption{The X-ray absorption column density $N_H$ as a function of the redshift, combining the results of our sources (filled rhombs) and the ones of a larger sample (plus and cross signs) analyzed in the soft X-ray energy range.}
\end{figure}

The $N_H$ distribution shown in Figure 2 appears to be asymmetric,
with a peak located at about $log(N_H)=21$ and tails extending
towards low and high values up to $log(N_H)=20$ and $log(N_H)=23$,
respectively. It is worth to note that the statistical
significance of this result is strongly affected by the limited
number of objects in our sample. Therefore we compare our sample
with the larger sub-sample from the literature observed in the
$0.2-10$ keV energy range from XMM-Newton that includes 16 RL
objects identified as blazars and with a specific value of the
absorption. We underline that this one includes only one object
belonging to our sample (the blazar QSO B0836+710). The $N_H$
distribution of the larger sample is shown in Figure 3, with a
peak located at $log(N_H)=22$ and a tail extending towards lower
values up to $log(N_H)=20$. Comparing the distributions in Figure
3 and in Figure 2, we note that sources with absorbing column
density around $log(N_H)=23$ are "missing". This evidence is
expected in view of the fact that in the soft X-ray energy range
($0.2-10$ keV) the sources with an absorbing column density $N_H
\ge 1.3$ $10^{23}$ cm$^{-2}$ are almost completely obscured.

This occurrence strongly suggests that data above $10$ keV
represent a crucial tool to construct an unbiased selection of
heavily absorbed objects.


The $N_H$-redshift trend of our sample is shown in Figure 4.

The comparison of this $N_H$-$z$ plot with those previously
analyzed by \citet{b54} at various redshifts for a sample of 32 RL
sources, seems to indicate that the property of the X-ray
absorbing gas evolves with cosmic time. This trend is shown in
Figure 5, where we added 3 objects analyzed by \citet{b17}. In
fact, at first glance, a correlation between $N_H$ and redshift
seems to exist and, compared to low redshifts ($z<2$), a shift
toward higher values of $N_H$ seems to be indicated at high
redshifts.

The sample of 32 sources includes 3 objects of our sample (QSO
B0836+710, PKS 0537-286 and PKS 2149-307)
and the $N_H$ values for these 3 sources as obtained through our
spectral analysis are in agreement with those presented in the
work of Yuan and collaborators (2006).

The existence of a correlation is also confirmed by the
statistical analysis: we obtain a correlation coefficient ($R$)
around 0.7 (Figure 5), when the results corresponding to our
sample are added to the sources investigated by \citet{b37},
\citet{b17} and \citet{b54}.

All together these results seem to support the hypothesis of an
intrinsic absorber. Three further hints support its intrinsic
nature:

\begin{itemize}

\item{the existence of a spatial isotropy of the inter-galactic
absorbers is most unlikely;}

\item{the X-ray absorption is associated with RL but not with RQ
objects \citep{b55}. This evidence supports the hypothesis that
the absorber are physically associated with the RL quasars;}

\item{the change of the $N_H$ distribution that seems to occur at
$z\sim2$ and the decrease of the slope in the $N_H$-$z$ relation
at $z>2$ could indicate an effect of cosmic evolution. This effect
could be due, e.g., to a rate of stellar formation. In fact, an
increase of the stellar formation at $z\le2$ could reduce the
amount of gas in the host galaxy.}

\end{itemize}


\subsubsection{PHOTON-INDEX DISTRIBUTION}

The photon-index distribution was then investigated by adding the
FSRQ 3C 273 and 4C 04.42 in our analysis ($\Gamma\sim1.6$ and
$\sim1.3$, respectively, as obtained by our broad-band analysis).
The $\Gamma$ distribution for the complete sample is shown in
Figure 6 and the mean value obtained in $0.2-100$ keV energy band
is $\Gamma=1.39\pm0.01$.


\begin{figure}
\includegraphics[width=64mm]{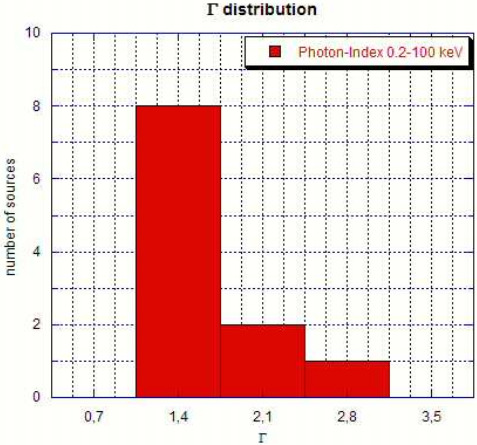}
 \caption{Distribution of the power-law photon-index for our sample in the $0.2-100$ keV energy range.}
\end{figure}

\begin{figure}
\includegraphics[width=64mm]{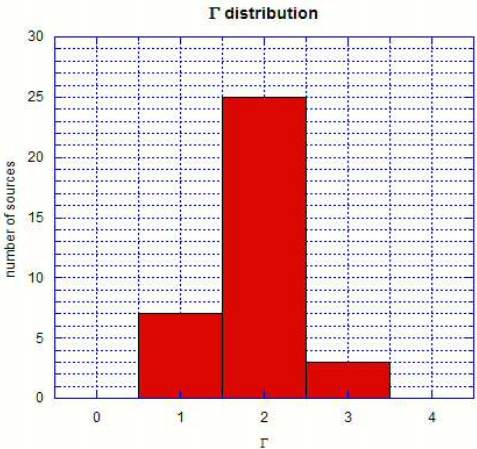}
 \caption{Distribution of the
photon-index resulting by the analysis performed in the works of
\citet{b37}, \citet{b17} and \citet{b54} and including 35 RL objects.}
\end{figure}

For comparison, in Figure 7 we also show the photon index
distribution derived from \citet{b37}, \citet{b17} and \citet{b54}
for 35 sources. The $\Gamma$ distribution of our sample shows a
peak at $\Gamma\sim1.4$, whereas the $\Gamma$ distribution of the
RL objects
shows a peak corresponding to a steeper value of $\Gamma$ $\sim2$.
However, we note here that our sample is mainly composed by FSRQ,
while the larger sample contains different types of RL objects -
not all identified as blazars. Therefore, in order to perform a
meaningful comparison, we have studied the photon index
distribution for sources -- in the larger sample -- identified as
blazars (19 objects). The peak of the $\Gamma$ distribution (see
Figure 8) is, again, around $2$.


\begin{figure}
\includegraphics[width=64mm]{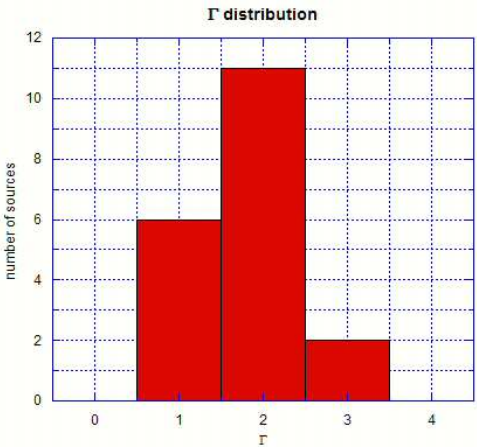}
 \caption{Photon-index distribution for 19 sources identified as blazars.}
\end{figure}

\begin{figure}
\includegraphics[width=64mm]{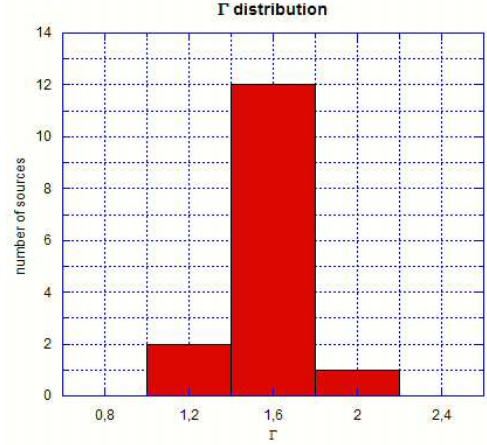}
 \caption{Photon-index distribution for 15 sources identified as FSRQ.}
\end{figure}

To account for the fact that most of the sources in our sample are
FSRQ-type objects and that FSRQs in X-rays show a lower value of
$\Gamma$ with respect to BL Lac objects, we studied the $\Gamma$
distribution including only the FSRQ sources (15 objects) in the
larger sample. In this case, the $\Gamma$ distribution in Figure 9
shows a peak at $\Gamma\sim1.6$, a value steeper than that for our
blazars sample.

The differences between the values of $\Gamma$ of our sample and
those of the larger sample could be due to the hard X-ray
selection of our INTEGRAL/IBIS sample, which is clearly biased
towards the flatter values of the photon index.

\section{Summary and conclusions}

In this work we presented a broad-band X-ray spectral study of a
sample of 9 blazars (with a redshift range of $0.1<z<3.7$)
observed with INTEGRAL, XMM-Newton and Swift. The main results can
be summarized as follows:

\begin{itemize}

\item{The broad-band spectra of all selected sources are well
reproduced with a power-law model absorbed by an amount of gas in
excess to the Galactic one ($N_H\sim10^{20}$-$10^{23}$cm$^{-2}$ in
rest-frame of the source; only an upper limit of $N_H\sim5$
$10^{20}$cm$^{-2}$ has been derived for the FSRQ PKS 2149-307).}



\item{The absorption seems to be a signature of a cold intrinsic
absorber, confirming and extending to larger sample previous
results quoted in the literature \citep{b8,b37,b54}.}


\item{The present work provides a further confirmation of the
existence of a $N_H$-redshift trend, obtained for a large sample
of RL objects (not only blazars).}

\item{The broad-band analysis of our sample of blazars revealed a
harder spectrum with a photon-index of the order of
$\Gamma\sim1.4$, compared to the value obtained with the
distribution including the FSRQ sources of the larger sample (15
objects).
Such a difference could be due to the hard X-ray selection
of our INTEGRAL/IBIS sample which is clearly biased towards
flatter values of the photon index.}


\item{We have found no evidence of reflection components
(reflection "hump" and iron emission line), with the exception of
the source IGR J22517+2218 that shows the presence of a weak iron
line. This result is expected in blazar-type objects.}




\end{itemize}

In conclusion, the analysis presented here has shown that our
INTEGRAL-sample selection favours objects heavily absorbed and
with a flatter value of spectral index. On the other hand, the
present broad-band analysis of INTEGRAL/IBIS, XMM-Newton and Swift
observations of RL QSOs confirms that the observed flattening is
common in these objects, and is clearly detected in eight quasars
of our sample (8/9). The assumption of an intrinsic origin and a
cold nature for the absorber is consistent with previous results
up to $10$ keV obtained by \citet{b37} and \citet{b54} with
XMM-Newton data only. However, a broken power-law model, as an
alternative explanation for the deficit of soft photons observed
in the majority of our sources, cannot be ruled out by the data.

\section*{Acknowledgments}

Authors acknowledge support from INAF and ASI via contract
I/008/07.

\end{document}